# Symmetry-Dependent Polarity Reversal of Bulk Spin-Orbit Torque in Single-Layer MnCoGa

Rongkun Han[1, 2†*], Mingsong Zhang[3†], Dahai Wei[1], Yuan Lu[2], Wenxu Zhang[3*], and Jianhua Zhao[1,4*]

[1]*State Key Laboratory of Semiconductor Physics and Chip Technologies, Institute of Semiconductors, Chinese Academy of Sciences, Beijing 100083, China*

[2]*Institut Jean Lamour, Université de Lorraine, CNRS, UMR 7198, Nancy, France*

[3]*School of Integrated Circuits Science and Engineering, University of Electronic Science and Technology of China, Chengdu 611731, China*

[4]*State Key Laboratory of Spintronics, Hangzhou International Innovation Institute, Beihang University, Hangzhou 311115, China*

**Abstract:** Controlling the magnitude and polarity of spin-orbit torque (SOT) is essential for manipulating magnetization in spintronic devices. Here, we realize bulk spin-orbit torque (SOT) in a single-layer $Mn_{1.6}Co_{1.4}Ga$ (MCG) Heusler alloy and achieve lattice strain-controlled reversal of SOT polarity. SOT-driven magnetization switching is achieved over a thickness range of 3-20 nm, with the switching polarity reversed between the strongly strained ultrathin regime and thicker films. First-principles calculations reveal a sizable intrinsic spin Hall conductivity (SHC) in MCG originating from its topological band structure, which is enhanced by the tetragonal distortion without changing sign. Together with the thickness-dependent structural evolution, we demonstrate that the reversal of SOT polarity does not originate from a sign change of the SHC, but rather from lattice distortion induced symmetry breaking that modifies the conversion of spin current into a net bulk SOT, thereby controlling its polarity. These results establish strain-controlled structural symmetry as an additional degree of freedom for tuning bulk SOT in magnetic single layers.

**Key words**: Heusler alloy; spin-orbit torque; spin Hall conductivity; first-principles calculations

[†]These authors contributed equally to this work.

[*]Corresponding authors. Email: jianhuazhao@buaa.edu.cn, xwzhang@uestc.edu.cn, and rongkun.han@univ-lorraine.fr

## Introduction

Controlling the magnitude and polarity of spin-orbit torque (SOT) is a central challenge for electrically manipulating magnetization in spintronic devices [1,2]. In conventional heavy metal/ferromagnet (HM/FM) bilayers, the polarity of SOT is determined by the sign of the spin Hall effect (SHE) in the HM, while interfacial spin transmission limits the SOT efficiency and the thickness of the magnetic layer [3-6]. Magnetic single layers with an intrinsic SHE generates bulk SOT without an additional spin source, avoiding interfacial spin transmission losses and enabling thicker magnetic layers [7-11]. However, the realization of bulk SOT requires broken out-of-plane symmetry in the magnetic layer. For a magnetic single layer with mirror symmetry with respect to the *xy* plane [Fig. 1(a)], spin accumulations generated across the film thickness cancel, resulting in zero net SOT. Breaking global or local symmetry of the film, for example through compositional gradients [12,13] or non-centrosymmetric crystal structures [14,15], lifts the cancellation and generate a net bulk SOT [Fig. 1(b)] [16,17]. More importantly, structural symmetry also provides a degree of freedom for controlling the magnitude and polarity of bulk SOT. The polarity of bulk SOT is also sensitive to structural asymmetry, further highlighting the role of symmetry in determining the effective torque [18-20].

In magnetic single layers, bulk SOT depends on both intrinsic SHE and its conversion into torque. While large SHE is often associated with strong spin orbit coupling (SOC) in heavy element systems [21], Mn-Co-Ga Heusler alloys provide an alternative platform where sizable intrinsic SHE originates from nontrivial band structures [22-25]. In particular, the Weyl semimetal $Co_2MnGa$ exhibits a large SHE associated with Berry curvature near the Fermi level [26-28], while simultaneously possessing a high Curie temperature, large spin polarization, and tunable magnetic anisotropy. Moreover, the band structure and magnetic properties of $Co_2MnGa$ are highly sensitive to lattice distortion and strain engineering [29,30]. Increasing the Mn concentration transforms cubic $Co_2MnGa$ into tetragonal MnCoGa, where lattice distortion breaks the cubic symmetry and induces strong perpendicular magnetic anisotropy (PMA) [31-33]. Such lattice distortion modifies the electronic structure and intrinsic SHE while simultaneously changing the symmetry conditions. MnCoGa therefore provides a platform for investigating how structural symmetry determines the magnitude and polarity of bulk SOT.

In this study, we demonstrate bulk SOT-driven magnetization switching in a $Mn_{1.6}Co_{1.4}Ga$ (MCG) single layer with PMA induced by tetragonal lattice distortion. Current-induced magnetization switching is achieved over a thickness range from 3 to 20 nm, accompanied by a sign reversal of SOT polarity with increasing film thickness. First-principles calculations reveal that the intrinsic spin Hall conductivity (SHC) is enhanced by tetragonal lattice distortion while retaining the same sign. Combined with thickness-dependent structural characterization, we suggest that structural symmetry plays an important role in converting the spin current into a net bulk SOT. The bulk SOT is not determined by the SHE alone, but also by the symmetry-dependent conversion of spin current into torque. Our work establishes lattice structural symmetry as a general strategy for controlling the polarity and efficiency of bulk SOT.

## Experiment details

MnCoGa films were grown on GaAs (001) substrates by a molecular-beam epitaxy (MBE) system. Following the growth of a 200-nm GaAs buffer layer, $Mn_{1.6}Co_{1.4}Ga$ films with thicknesses ranging from 3 to 40 nm were deposited at 280 °C and capped with 3 nm Al to prevent oxidation [31]. The crystal structure and epitaxial relationship were characterized by X-ray diffraction (XRD) $\theta$-$2\theta$ and $\varphi$ scans. Magnetic properties were measured using a vibrating sample magnetometer, and transport measurements were performed in a physical property measurement system. The electronic band structures were investigated through density functional theory (DFT) with the full-potential local-orbital code (FPLO) [34,35]. The SHC was obtained from a Wannier based tight binding Hamiltonian using linear response calculations [36].

## Results and Discussion

### Thickness-Dependent Lattice Distortion and Magnetic Anisotropy

Bulk $Co_2MnGa$ crystallizes in a cubic structure with lattice constants *a*=*b*=*c*=5.77 Å. Increasing the Mn concentration can expand the lattice and introduce a lattice mismatch with the GaAs substrate [32,33], leading to in-plane compressive strain and out-of-plane elongation of the *c*-axis. XRD $\theta$-$2\theta$ patterns display clear MCG (002) and (004) diffraction peaks, confirming highly oriented growth [Fig. 1(c)]. The lattice constant *c* significantly increases to 6.05-6.21 Å, indicating a tetragonal distortion. The lattice distortion exhibits a strong thickness dependence. For films between 3 and 10 nm, the *c*-axis increases from 6.05 to 6.21 Å, indicating progressive strain accumulation during epitaxial growth. As the thickness exceeds 10 nm, the *c*-axis gradually decreases and reaches 6.13 Å at 40 nm [Fig. 1(f)], reflecting strain relaxation. The epitaxial relationship MCG(001)[110]∥GaAs(001)[110] is confirmed by $\varphi$ scans, as shown in Fig. 1(d), and the presence of both (002) and (111) reflections indicates a high crystalline order.

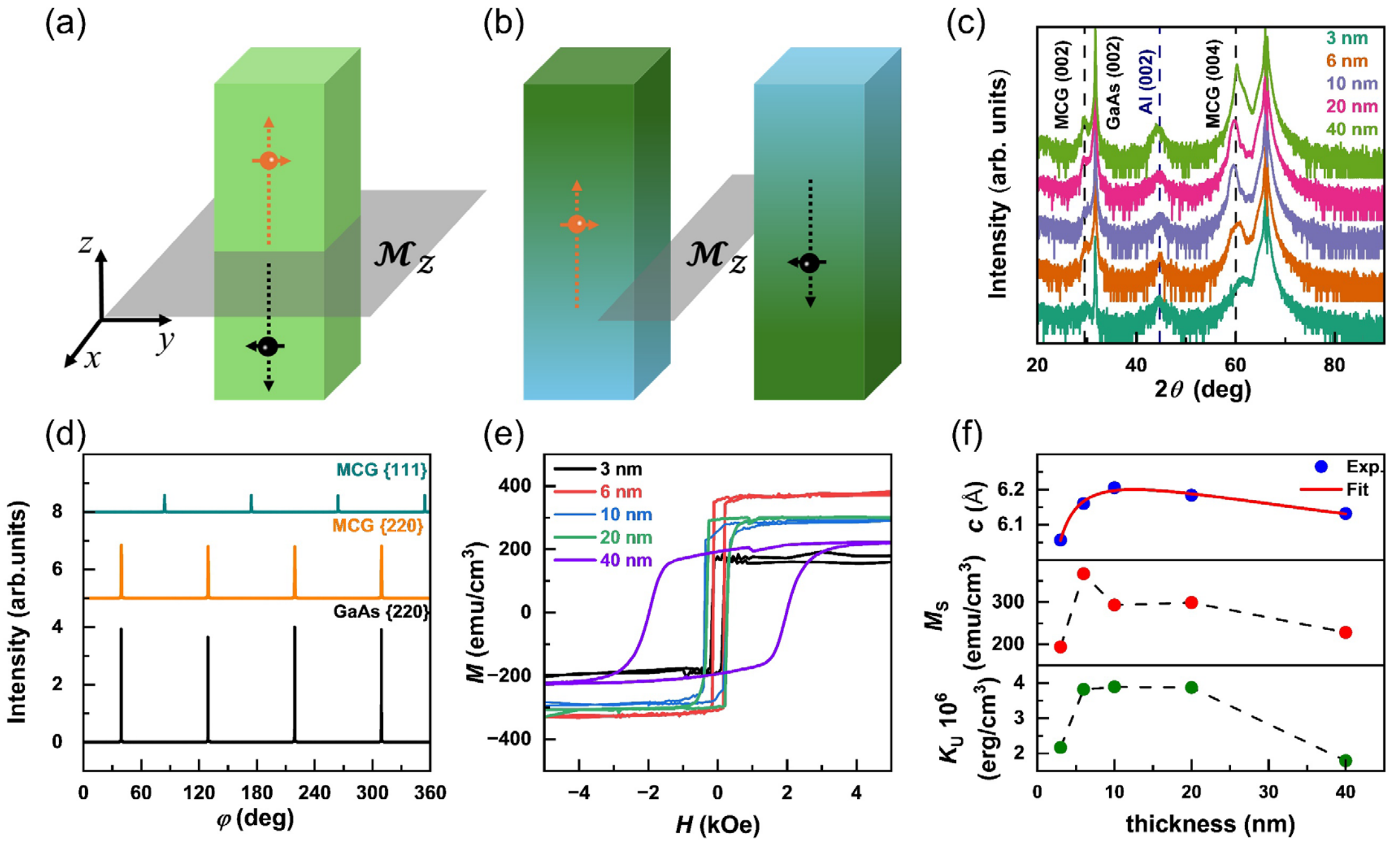

Fig. 1. (a) The spin current in a magnetic single layer with $\mathcal{M}_z$ symmetry. (b) The spin current in a magnetic single layer with broken $\mathcal{M}_z$ symmetry. (c) The XRD $\theta$-$2\theta$ scan results of MCG single layer with different thickness. (d) The XRD $\varphi$ scan of MCG {111} planes, MCG {220} planes, and GaAs {220} planes. (e) The *M*-*H* curves of MCG single layer at 300 K with different thickness. (f) The *c*, $M_S$, and $K_U$ of MCG as a function of thickness at 300 K.

Out-of-plane hysteresis loops at 300 K reveal robust PMA in MCG films across the entire thickness range, as shown in Fig. 1(e). Combining the XRD results, the emergence of PMA is attributed to tetragonal lattice distortion that lowers the cubic symmetry. The saturation magnetization ($M_s$) of MCG films ranges from 194 to 367 emu/cm$^3$ [Fig. 1(f)]. The perpendicular anisotropy energy $K_U$ is evaluated for different film thicknesses (see S1 in the Supporting Information). The $K_U$ reaches 3.8 ×10$^6$ erg/cm$^3$ for films with thicknesses between 6 and 20 nm and decreases to 1.8 ×10$^6$ erg/cm$^3$ at 40 nm [Fig. 1(f)]. This reduction correlates with strain relaxation, where the gradual release of in-plane compressive strain reduces the *c*-axis and consequently weakens the PMA.

**Reversal of bulk spin-orbit torque polarity**

Current-driven magnetization switching was measured by applying a sequence of 50 $\mu$s pulsed currents ($I_{pulse}$) under an in-plane magnetic field ($H_x$), as illustrated in Fig. 2(a). As shown in Fig. 2(b), the anomalous Hall effect (AHE) curves confirm robust PMA in MCG films. Figure 2(c) presents the AHE resistance $R_{xy}$ as a function of $I_{pulse}$ for MCG single layer with different thickness under $H_x$=+100 Oe. Deterministic magnetization switching occurs for thicknesses from 3 to 20 nm, while no switching is observed in the 40 nm film. Furthermore, the 3-nm-thick MCG film exhibits an opposite switching polarity compared with that of thicker films under the same in-plane field. Figures 2(d) and 2(e) show the magnetization switching curves for MCG films with thicknesses of 3 and 10 nm measured under different $H_x$. The 3-nm-thick MCG film shows a switching polarity that is reversed compared with the thicker films under the same $H_x$ (see S2 in the Supporting Information), consistent with the behavior in Fig. 2(c). This observation indicates that the sign of the effective SOT depends strongly on film thickness. Figure 2(f) summarizes the switching ratio (SR) and the critical current density $J_c$ as a function of film thickness. The 3-nm MCG film exhibits the highest switching ratio of 62% and the largest $J_c$=3.18×10$^7$ A/cm$^2$. As the film thickness increases, the $J_c$ decreases monotonically and reaches 1.65×10$^7$ A/cm$^2$ at 20 nm.

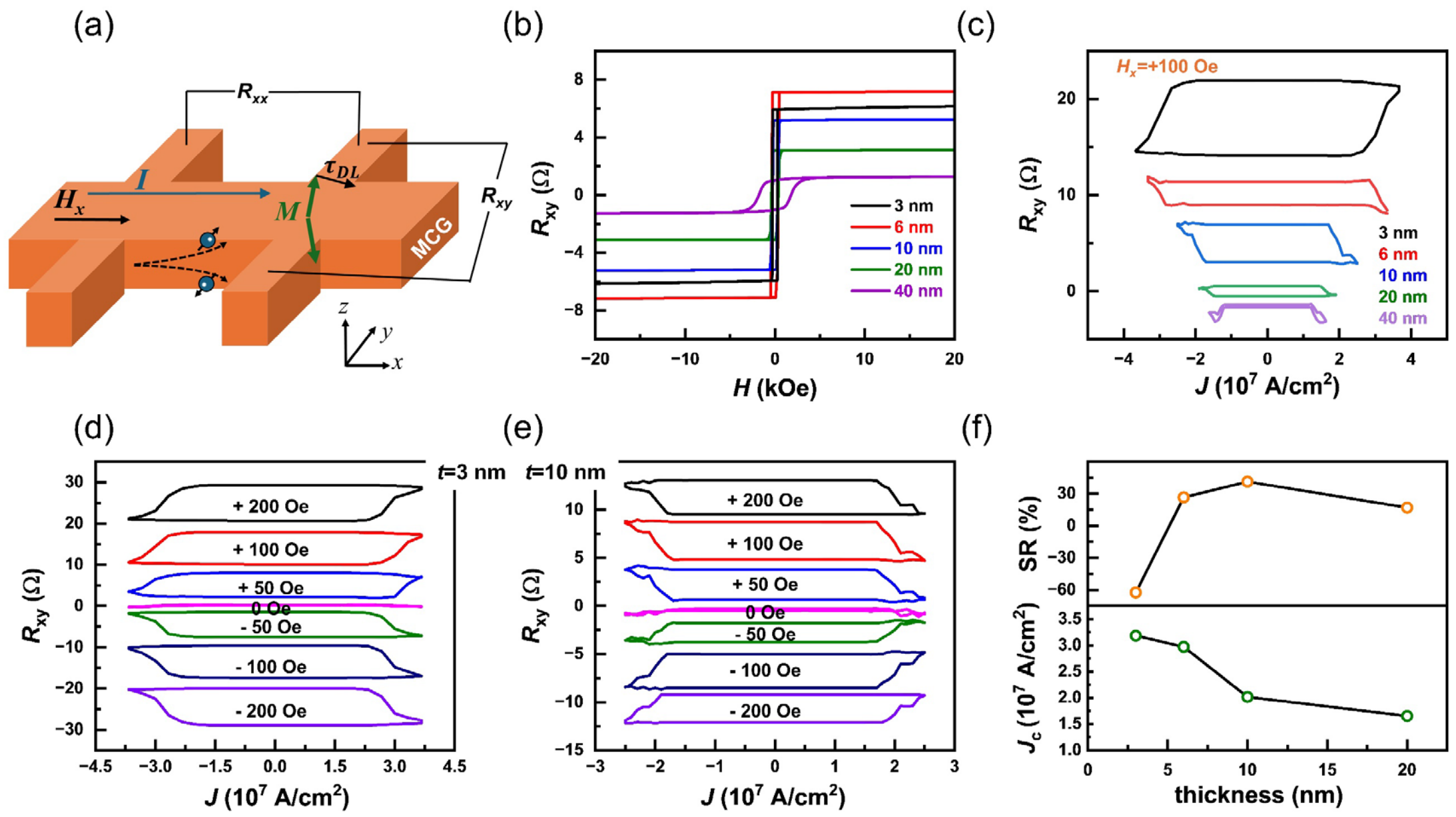

Fig. 2. (a) Schematics of the electrical transport measurement in MCG single layer Hall devices. (b) The AHE curves of MCG single layer at 300 K with different thickness. (c) The current-driven magnetization switching curves of MCG single layer at +100 Oe with different thickness. The magnetization switching curves of MCG with 3 nm (d) and 10 nm (e) at different in-plane magnetic field. (f) The magnetization switching ratio (SR) and critical current density as a function of thickness of MCG.

Second harmonic Hall voltage measurements were performed to quantify the SOT in MCG films with different thicknesses [Fig. 3(a)]. The first-harmonic Hall voltage $V_{1\omega}$ exhibits a parabolic dependence in the low field range, while the second-harmonic Hall voltage $V_{2\omega}^{x(y)}$ varies linearly with the in-plane field $H_x(H_y)$, as shown in Fig. 3(b). The damping-like and field-like effective field $H_{DL}$ and $H_{FL}$ can be extracted using [37]:

$$H_{DL(FL)} = -2\frac{\partial V_{2\omega}^{x(y)}}{\partial H_{x(y)}} \Big/ \frac{\partial^2 V_{1\omega}^{x(y)}}{\partial H_{x(y)}^2}. \quad (1)$$

By linear fitting the field dependent $V_{2\omega}$, the damping-like torque efficiency $\chi_{DL}$ is obtained as $\chi_{DL} = H_{DL}/J$, where $J$ is current density. The calculated $\chi_{DL}$ in MCG single layers is larger than the $\chi_{FL}$, indicating that the SOT is dominated by the DL component. A clear sign reversal of the $\chi$DL is observed as a function of thickness. The 3 nm film exhibits a negative $\chi_{DL}$ of $-22.71\times10^{-7}$ Oe·cm$^2$/A, while $\chi_{DL}$ becomes positive for thicker films and increases from 9.82 to $18.23\times10^{-7}$ Oe·cm$^2$/A as the thickness grows from 6 to 20 nm [Figs 3(c) and 3(d)]. This behavior contrasts with the typical $1/t$ of SOT efficiency observed in heavy metal/Co-based Heusler alloys heterostructures[33] and suggests a bulk SOT mechanism.

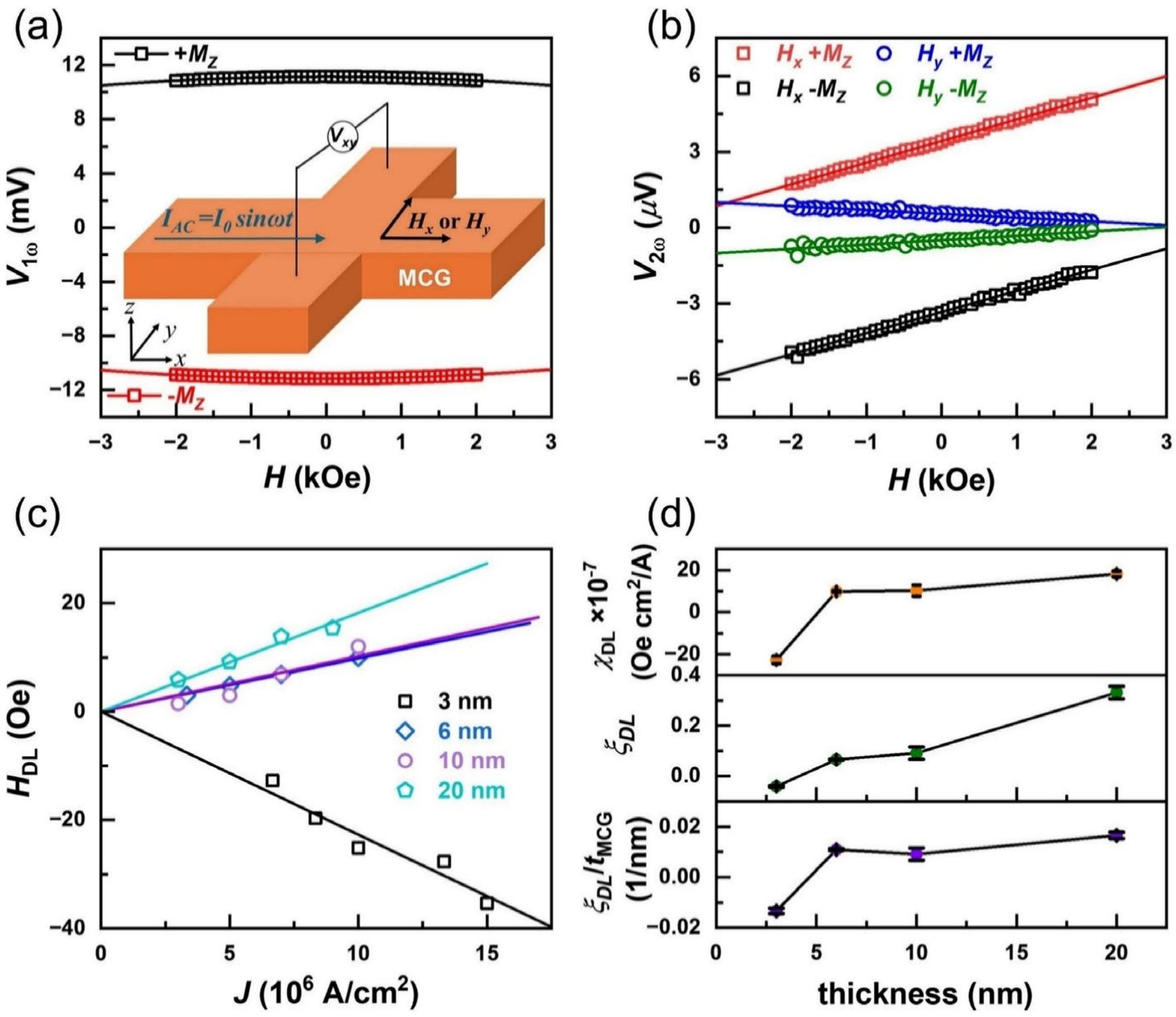

Fig. 3. (a) Variation of first order harmonic signal $V_{1\omega}$ with $H_x$ in MCG (20 nm) single layer. Inset in the schematic diagram of HHVR measurement. (b) Variations of second order harmonic signal $V_{2\omega}$ with $H_x$ and $H_y$, respectively. (c) $H_{DL}$ as a function of applied $J$ of MCG single layer with different thickness. (d) The $\chi_{DL}$, $\xi_{DL}$, and $\xi_{DL}/t_{MCG}$ as a function of thickness of MCG.

The DL torque efficiency $\xi_{DL}$ is calculated as:

$$\xi_{DL}=\frac{2eM_s t_{MCG}}{\hbar}\frac{H_{DL}}{J}, \qquad (2)$$

where $e$ is the electron charge and $\hbar$ is the reduced Planck constant. As shown in Fig. 3(d), $\xi_{DL}$ increases with film thickness, rising from -0.042 at $t_{MCG}$=3 nm to 0.303 at $t_{MCG}$=20 nm. The thickness normalized $\xi_{DL}/t_{MCG}$ shows a similar trend, increasing from -0.01339/nm at 3 nm to 0.01657/nm at 20 nm. This behavior, together with the thickness dependence of $\chi_{DL}$, supports a bulk origin of the SOT in MCG single layers.

In magnetic single layers, bulk SOT generally requires breaking the out-of-plane mirror symmetry. Several mechanisms can induce such symmetry breaking, including vertical composition gradient, bulk Dzyaloshinskii–Moriya interaction (DMI), and out-of-plane spin current ($\sigma_z$). To identify the dominant mechanism in MCG, we systematically examine these possibilities. No threshold-like variation characteristic of DMI is observed, indicating that the bulk SOT does not arise from this mechanism (see S3 in Supporting Information). Moreover, the out-of-plane effective field $H_{oop}$ showed no systematic dependence on current density, suggesting a negligible contribution from $\sigma_z$ (see S4 in Supporting Information). Compositional gradients are also unlikely to account for the observed behavior. The SOT in MCG exhibits thickness dependence, including a sign reversal and suppression at large thickness, which is inconsistent with composition driven mechanisms. Furthermore, reversing the Mn concentration gradient does not change the switching polarity or amplitude, ruling out composition induced bulk SOT (see S5 in Supporting Information).

The observed behavior in MCG films correlates directly with the thickness dependent lattice distortion. The out-of-plane lattice constant $c$, $\xi_{DL}$, and $\xi_{DL}/t_{MCG}$ all show thickness dependence, as summarized in Figs. 1(f) and 3(d). For ultrathin films (3 nm), the structural disorder is stronger, and the SOT efficiency is relatively large but negative. As the thickness increases to 6-20 nm, the crystalline ordering improves, accompanied by a reversal of the SOT sign to positive. When the thickness further increases to 40 nm, lattice relaxation significantly reduces the distortion and suppresses the bulk SOT. These results suggest that the bulk SOT in MCG is related to lattice constant and distortion, which breaks the structural symmetry along the $z$ direction.

**Lattice distortion dependent intrinsic spin Hall conductivity**

Given the correlation between bulk SOT and intrinsic SHE[7-9], we performed first-principles calculations to examine the origin of the bulk SOT in MCG by investigating the effect of lattice distortion on its SHC (see S6 in Supporting Information for calculation details). A supercell structure based on the $Co_2MnGa$ was constructed by replacing 25% of the Co atoms with Mn atoms to obtain $Mn_{1.5}Co_{1.5}Ga$ [Fig. 4(a)]. The initial lattice constant is $a$=$b$=$c$=5.77 Å, and the $c$-axis length is extended to 6.05 Å, 6.15 Å, and 6.21 Å, respectively, as the tetragonal lattice distortion increases. As shown in Fig. 4(b), the calculated $M_S$ falls within the experimental range shown in Fig. 1(f). The calculated $M_S$ increases from 150 emu/cm$^3$ in the undistorted structure to 264 emu/cm$^3$ at $c$=6.05 Å. This discrepancy likely arises from strong structural disorder in the ultrathin film, which suppresses the effective magnetic moment. With further lattice distortion, the calculated $M_S$ increases to 269 emu/cm$^3$ at $c$=6.21 Å, which is approximately 8% lower than the experimental value. The calculated $K_U$ exhibits a much stronger dependence on lattice distortion. At $c$=6.05 Å, $K_U$ reaches $1.06\times10^6$ erg/cm$^3$, a value more than an order of magnitude larger than that of the undistorted structure, and further increases to $3.13\times10^6$ erg/cm$^3$ at $c$=6.21 Å. Both the magnitude and the evolution of the calculated $K_U$ agree well with the experimental observations. The calculated SHC increases from 238.48 $\Omega^{-1}$cm$^{-1}$ in the cubic structure to 491.97 $\Omega^{-1}$cm$^{-1}$ at $c$=6.05 Å and further to 655.53 $\Omega^{-1}$cm$^{-1}$ at $c$=6.21 Å, demonstrating a strong dependence on lattice distortion. Calculations show that the tetragonal distortion modifies the electronic structure of $Mn_{1.5}Co_{1.5}Ga$. The majority-spin $t_{2g}$ and $e_g$ bands gradually shift toward lower energies, while the minority-spin bands move upward (see S7 in Supporting Information). Consequently, the Fermi level progressively intersects the $t_{2g}$ and $e_g$ manifolds at the high-symmetry points L and X, as highlighted by the white dashed circles in Figs. 4(c) to 4(f). Previous studies have shown that positioning the Fermi level within the $t_{2g}$ and $e_g$ manifolds in Heusler compounds favors a large SHC [38]. In addition, tetragonal distortion introduces several Weyl points near the Fermi level (see S7 in Supporting Information), which further contributes to the Berry curvature. The combined effects of band structure reconstruction and the emergence of Weyl points enhance the SHC with increasing tetragonal distortion.

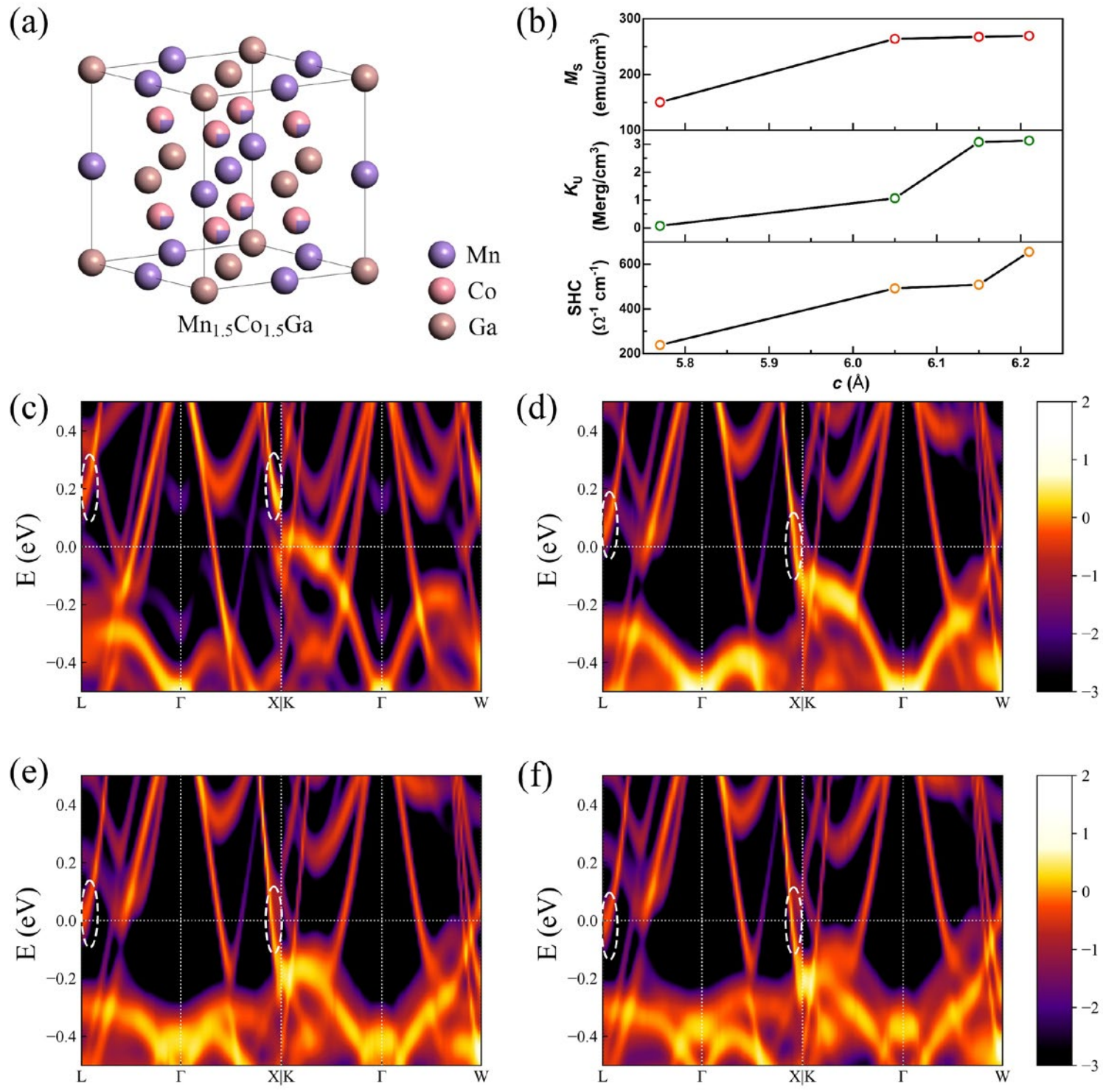

Fig. 4. (a) The crystal structure of $Mn_{1.5}Co_{1.5}Ga$. (b) The dependence of the calculated MS, KU, and SHC on lattice distortion. The band structures projected onto $t_{2g}$ and $e_g$ orbitals of $Mn_{1.5}Co_{1.5}Ga$ with $c$=5.77 Å (c), 6.05 Å (d), 6.15 Å (e), and 6.21 Å (f), respectively. The color bar represents the spectral weights on a logarithmic scale, and the white dashed circles serve as visual guides.

**Symmetry-dependent conversion from spin current to SOT**

The calculated SHC increases monotonically with tetragonal lattice distortion while remaining the same sign throughout. This behavior contrasts with the experimentally observed reversal of the SOT polarity, indicating that the sign of the effective torque cannot be determined by the intrinsic SHE alone. As discussed above, mirror symmetry with respect to the *xy* plane enforces a cancellation of the SOT, even in the presence of a large intrinsic SHC. A vertical structural asymmetry lifts this compensation and allows net bulk SOT. Therefore, the polarity and magnitude of the SOT also depend on how spin current is converted into torque under broken structural symmetry.

To gain further insight into the thickness dependence of the SOT, we consider the evolution of lattice distortion during epitaxial growth. The average *c*-axis lattice firstly increases and then decreases with film thickness, indicating a competition between strain accumulation in the initial growth stage and strain relaxation in thicker films. We describe this behavior using a phenomenological lattice distortion profile,

$$c(z) = c_0\left(1 - e^{-z/t_s}\right)e^{-z/t_r} + c_\infty, \quad (3)$$

where $t_s$ and $t_r$ represent the characteristic lengths for strain build-up and strain relaxation, respectively. $c_0$ is the distortion amplitude, and $c_\infty$ is the fully relaxed lattice constant. Since XRD measures the averaged lattice constant, the measured value is

$$\bar{c}(t) = \frac{1}{t}\int_0^t c(z)\,dz = \frac{c_0}{t}\left[t_r(1 - e^{-\frac{t}{t_r}}) - \frac{t_r t_s}{t_r + t_s}(1 - e^{-\frac{t}{t_s}-\frac{t}{t_r}})\right] + c_\infty, \quad (4)$$

which is used to fit the experimental data. As shown in Fig. 1(f), the fitting yields $t_s \approx 1.61$ nm and $t_r \approx 41.4$ nm, indicating that lattice distortion is established within the first few nanometers of growth, whereas strain relaxation becomes dominant at much larger thicknesses.

Because the intrinsic SHC increases monotonically with lattice distortion, the spatial distribution of the *c* axis lattice constant determines the spatial distribution of spin current generation across the film thickness. The corresponding lattice distortion gradient is

$$\frac{\partial c(z)}{\partial z} = c_0\left(-\frac{1}{t_r}e^{-\frac{z}{t_r}} + \frac{t_r + t_s}{t_r t_s}e^{-\frac{t}{t_s}-\frac{t}{t_r}}\right). \quad (5)$$

The calculated gradient is positive for the 3 nm film but becomes negative for thicker films, indicating a reversal of the structural asymmetry across the film thickness. A positive gradient corresponds to stronger lattice distortion near the film surface, whereas a negative gradient indicates that the larger lattice distortion shifts toward the substrate interface. Consequently, the dominant contribution to spin current generation changes from the top region of the film to the bottom region, leading to a reversal of the effective bulk SOT polarity. This evolution is consistent with the experimentally observed sign reversal of the SOT. These results identify the lattice distortion gradient as the key parameter governing the symmetry dependent conversion of spin current into torque and provide a physical picture for the thickness dependent reversal of bulk SOT.

## Conclusion

In conclusion, we demonstrate the tetragonal lattice distortion enables bulk SOT-driven magnetization switching in MCG single layer with different thicknesses. The switching polarity and SOT efficiency exhibit a strong thickness dependence, including a sign reversal as the film thickness increases, whereas the calculated intrinsic SHC is enhanced by lattice distortion without changing sign. This contrasting behavior shows that the polarity of SOT is not only determined by the intrinsic SHE, but also by its symmetry-dependent conversion into torque. These results establish structural symmetry as an additional control parameter for manipulating the polarity and efficiency of bulk SOT in magnetic single layers.

**ACKNOWLEDGMENTS**

The work was supported by the research Start-up Funds of Hangzhou International Innovation Institute of Beihang University No. 2024KQ01 and National Natural Science Foundation of China (Grant No. 12274060). Y. L. acknowledges the support by the French National Research Agency (ANR) SOT spin LED project (no. ANR-22-CE24-0006-01), CHIFTS project (No. ANR-23-CE09-0006-003), and PEPR-SPIN OptoSpinCom project (No. ANR-24-EXSP-0010).

**Reference:**

[1] X. Qiu, K. Narayanapillai, Y. Wu, P. Deorani, D.-H. Yang, W.-S. Noh, J.-H. Park, K.-J. Lee, H.-W. Lee, and H. Yang, Spin–orbit-torque engineering via oxygen manipulation, Nat. Nanotechnol. 10, 333 (2015).
[2] D. MacNeill, G. Stiehl, M. Guimaraes, R. Buhrman, J. Park, and D. Ralph, Control of spin–orbit torques through crystal symmetry in WTe2/ferromagnet bilayers, Nat. Phys. 13, 300 (2017).
[3] L. Liu, O. Lee, T. Gudmundsen, D. Ralph, and R. Buhrman, Current-Induced Switching of Perpendicularly Magnetized Magnetic Layers Using Spin Torque from the Spin Hall Effect, Phys. Rev. Lett. 109, 096602 (2012).
[4] C. Zhang, S. Fukami, H. Sato, F. Matsukura, and H. Ohno, Spin-orbit torque induced magnetization switching in nano-scale Ta/CoFeB/MgO, Appl. Phys. Lett. 107, 012401 (2015).
[5] H.-Y. Lee, S. Kim, J.-Y. Park, Y.-W. Oh, S.-Y. Park, W. Ham, Y. Kotani, T. Nakamura, M. Suzuki, and T. Ono, Enhanced spin–orbit torque via interface engineering in Pt/CoFeB/MgO heterostructures, APL Mater. 7, 031110 (2019).
[6] M. A. Wahada, E. Şaşıoğlu, W. Hoppe, X. Zhou, H. Deniz, R. Rouzegar, T. Kampfrath, I. Mertig, S. S. Parkin, and G. Woltersdorf, Atomic scale control of spin current transmission at interfaces, Nano Lett. 22, 3539 (2022).
[7] L. Liu, J. Yu, R. González-Hernández, C. Li, J. Deng, W. Lin, C. Zhou, T. Zhou, J. Zhou, and H. Wang, Electrical switching of perpendicular magnetization in a single ferromagnetic layer, Phys. Rev. B 101, 220402 (2020).
[8] R. Zhang, L. Liao, X. Chen, T. Xu, L. Cai, M. Guo, H. Bai, L. Sun, F. Xue, and J. Su, Current-induced magnetization switching in a CoTb amorphous single layer, Phys. Rev. B 101, 214418 (2020).
[9] K. Dong, C. Sun, L. Zhu, Y. Jiao, Y. Tao, X. Hu, R. Li, S. Zhang, Z. Guo, and S. Luo, Current-Induced Magnetic Switching in an L10 FePt Single Layer with Large Perpendicular Anisotropy Through Spin–Orbit Torque, Engineering 12, 55 (2022).
[10] X. R. Wang, Anomalous spin Hall and inverse spin Hall effects in magnetic systems, Commun. Phys. 4, 55 (2021).
[11] Q. Yang, D. Han, S. Zhao, J. Kang, F. Wang, S.-C. Lee, J. Lei, K.-J. Lee, B.-G. Park, and H. Yang, Field-free spin–orbit torque switching in ferromagnetic trilayers at sub-ns timescales, Nat. Commun. 15, 1814 (2024).
[12] Z. Zheng, Y. Zhang, V. Lopez-Dominguez, L. Sánchez-Tejerina, J. Shi, X. Feng, L. Chen, Z. Wang, Z. Zhang, and K. Zhang, Field-free spin-orbit torque-induced switching of perpendicular magnetization in a ferrimagnetic layer with a vertical composition gradient, Nat. Commun. 12, 4555 (2021).
[13] X. Xie, X. Zhao, Y. Dong, X. Qu, K. Zheng, X. Han, X. Han, Y. Fan, L. Bai, and Y. Chen, Controllable field-free switching of perpendicular magnetization through bulk spin-orbit torque in symmetry-broken ferromagnetic films, Nat. Commun. 12, 2473 (2021).
[14] P. Wadley, B. Howells, J. Železný, C. Andrews, V. Hills, R. P. Campion, V. Novák, K. Olejník, F. Maccherozzi, and S. Dhesi, Electrical switching of an antiferromagnet, Science 351, 587 (2016).

[15] M. Jiang, H. Asahara, S. Sato, T. Kanaki, H. Yamasaki, S. Ohya, and M. Tanaka, Efficient full spin–orbit torque switching in a single layer of a perpendicularly magnetized single-crystalline ferromagnet, Nat. Commun. 10, 2590 (2019).
[16] Z. Han, Y. Huo, Y. Yu, H. Wu, L. Gao, Z. Ma, G. Zhang, and M. Jiang, Unveiling the potential of spin–orbit torque in a magnetic single layer for advancing spintronics application, Appl. Phys. Rev. 13, 011305 (2026).
[17] Q. Liu, L. Zhu, X. S. Zhang, D. A. Muller, and D. C. Ralph, Giant bulk spin–orbit torque and efficient electrical switching in single ferrimagnetic FeTb layers with strong perpendicular magnetic anisotropy, Appl. Phys. Rev. 9, 021402 (2022).
[18] Z. Zheng, S. Shi, Z. Liu, Q. Zhang, N. A. Shahed, H. Mavani, G. Zhou, Q. Chen, C. Zhang, and Z. Zhou, Spin-Orbit Torque Induced by Switchable Crystal Inversion Symmetry Breaking, Adv. Mater., e73219 (2026).
[19] Y. Liu, H. Zheng, Q. Shen, Y. Ma, J. Liu, B. Rong, C. Man, L. Sun, H. Xie, and Y. Wu, Deterministic polarity control of self-generated spin–orbit torque via inversion symmetry engineering, Appl. Phys. Rev. 13 (2026).
[20] M. Tang, K. Shen, S. Xu, H. Yang, S. Hu, W. Lü, C. Li, M. Li, Z. Yuan, and S. J. Pennycook, Bulk spin torque-driven perpendicular magnetization switching in L10 FePt single layer, Adv. Mater. 32, 2002607 (2020).
[21] F. Zheng, M. Zhu, J. Dong, X. Li, Y. Zhou, K. Wu, and J. Zhang, Anatomy of the spin Hall effect in ferromagnetic metals, Phys. Rev. B 109, 224401 (2024).
[22] L. Leiva, S. Granville, Y. Zhang, S. Dushenko, E. Shigematsu, T. Shinjo, R. Ohshima, Y. Ando, and M. Shiraishi, Giant spin Hall angle in the Heusler alloy Weyl ferromagnet $Co_2MnGa$, Phys. Rev. B 103, L041114 (2021).
[23] M. Aoki, Y. Yin, S. Granville, Y. Zhang, N. V. Medhekar, L. Leiva, R. Ohshima, Y. Ando, and M. Shiraishi, Gigantic anisotropy of self-induced spin-orbit torque in Weyl ferromagnet $Co_2MnGa$, Nano Lett. 23, 6951 (2023).
[24] M. Wang, C. Pan, N. Xie, X. Qiu, Y. Li, L. Lang, S. Wang, D. Cheng, W. Fan, and S. M. Zhou, Large Spin Hall Efficiency and Current-Induced Magnetization Switching in Ferromagnetic Heusler Alloy $Co_2MnAl$-Based Magnetic Trilayers, Adv. Sci. 12, 2407171 (2025).
[25] Y. Ji, W. Zhang, H. Zhang, and W. Zhang, Spin Hall conductivity and anomalous Hall conductivity in full Heusler compounds, New J. Phys. 24, 053027 (2022).
[26] S. Ouardi, G. H. Fecher, C. Felser, and J. Kübler, Realization of spin gapless semiconductors: The Heusler compound Mn 2 CoAl, Phys. Rev. Lett. 110, 100401 (2013).
[27] G. Xu, L. Tang, P. Wang, S. Pan, Y. Bai, X. Xiang, S. Zhang, J. Zhu, and F. Xu, Magnetic-and Spin-Gapless-Related Transport Properties in Sputtered $Mn_2CoGa$ Thin Films, physica status solidi (RRL)–Rapid Research Letters 18, 2300499 (2024).
[28] N. Shahi, A. K. Jena, G. K. Shukla, V. Kumar, S. Rastogi, K. Dubey, I. Rajput, S. Baral, A. Lakhani, and S.-C. Lee, Antisite disorder and Berry curvature driven anomalous Hall effect in the spin gapless semiconducting $Mn_2CoAl$ Heusler compound, Phys. Rev. B 106, 245137 (2022).
[29] N. Schulz, G. Pantano, D. DeTellem, A. Chanda, E. M. Clements, M. McGuire, A. Markou, C. Felser, D. Arena, and J. Gayles, Unraveling the structural dependency of

Weyl nodes in Co 2 MnGa, Phys. Rev. B 110, 054419 (2024).
[30] A. Kumar, S. Chaudhary, and S. Chandra, Effect of point defects and lattice distortions on the structural, electronic, and magnetic properties of Co 2 MnAl Heusler alloy, Phys. Rev. Mater. 8, 034405 (2024).
[31] Z.-F. Yu, J. Lu, H.-L. Wang, X.-P. Zhao, D.-H. Wei, J.-L. Ma, S.-W. Mao, and J.-H. Zhao, Tunable Perpendicular Magnetic Anisotropy in Off-Stoichiometric Full-Heusler Alloy $Co_2MnAl$, Chin. Phys. Lett. 36, 067502 (2019).
[32] T. Kubota, D. Takano, Y. Kota, S. Mohanty, K. Ito, M. Matsuki, M. Hayashida, M. Sun, Y. Takeda, and Y. Saitoh, Magnetoelastic anisotropy in Heusler-type $Mn_{2-\delta}CoGa_{1+\delta}$ films, Phys. Rev. Mater. 6, 044405 (2022).
[33] H. Qin, X. Zhao, R. Han, H. Sun, Z. Xie, D. Wei, and J. Zhao, Efficient magnetization switching induced by spin–orbit torque in perpendicularly magnetized $Mn_{1+x}Co_{1-x}Al$ Heusler alloys, Appl. Phys. Lett. 124, 202406 (2024).
[34] K. Koepernik and H. Eschrig, Full-potential nonorthogonal local-orbital minimum-basis band-structure scheme, Phys. Rev. B 59, 1743 (1999).
[35] I. Opahle, K. Koepernik, and H. Eschrig, Full-potential band-structure calculation of iron pyrite, Phys. Rev. B 60, 14035 (1999).
[36] J. Zelezny, Wannier linear response, https://bitbucket.org/zeleznyj/wannier-linear-response/src/master/.
[37] M. Hayashi, J. Kim, M. Yamanouchi, and H. Ohno, Quantitative characterization of the spin-orbit torque using harmonic Hall voltage measurements, Phys. Rev. B 89, 144425 (2014).
[38] P. Jadaun, L. F. Register, and S. K. Banerjee, Rational design principles for giant spin Hall effect in 5d-transition metal oxides, Proceedings of the National Academy of Sciences 117, 11878 (2020).